\documentclass[12pt,a4paper]{article}
\usepackage[T1]{fontenc}
\usepackage{amssymb,amsfonts,amsmath,amsthm}
\usepackage[english]{babel}
\usepackage{longtable}
\usepackage[cp1251]{inputenc}
\usepackage{graphicx}
\usepackage{wrapfig}
\usepackage{hyperref}
\usepackage{authblk}
\usepackage{cite}

\hypersetup{
    colorlinks=true
}

\DeclareMathOperator{\rank}{\mathrm{rank}}

\DeclareMathOperator{\Diff}{\mathrm{Diff}}
\DeclareMathOperator{\ind}{\mathrm{ind}}
\DeclareMathOperator{\codim}{\mathrm{codim}}
\DeclareMathOperator{\ord}{\mathrm{ord}}

\newcommand{\Uk}[2]{\underset{#2}{#1}}
\newcommand{\w}[1]{w_{({#1})}}
\newcommand{\uu}[1]{u_{({#1})}}
\newcommand{\I}[1]{I_{({#1})}}
\newcommand{\cE}{{\cal E}}
\newcommand{\R}{\mathbb{R}}

\newcommand{\<}{\langle}
\renewcommand{\>}{\rangle}

\newcommand{\g}{\mathfrak{g}}
\newcommand{\h}{\mathfrak{h}}

\newcommand{\p}{\partial}

\newtheorem{proposition}{Proposition}
\newtheorem{theorem}{Theorem}

\theoremstyle{remark}
\newtheorem{remark}{Remark}

\title{Classification of scalar second-order differential equations with low-dimensional symmetry groups: The case of free action}

\author[1]{Alexey A. Magazev\thanks{magazev@omgtu.ru}}
\author[1]{Igor V. Shirokov\thanks{iv\_shirokov@mail.ru}}
\affil[1]{Omsk State Technical University, Omsk, Russia}
\date{}

\begin{document}

\maketitle

\begin{abstract}
Based on an original classification of differential equations by types of regular Lie group actions, we offer a systematic procedure for describing partial differential equations with prescribed symmetry groups. Using a new powerful algebraic technique based on the so-called covariant form of a differential equation, we give an effective algorithm for constructing differential equations whose symmetry groups regularly and freely act on the space of dependent and independent variables. As an application, we derive a complete classification of quasi-linear scalar second-order partial differential equations with regular free symmetry groups of dimension no greater than three.
\end{abstract}

\section*{Introduction}

Originally developed as a powerful tool for integrating ordinary differential equations, group analysis of differential equations has long been extended beyond this aim and extensively used in differential geometry and mathematical physics~\cite{Lie1880, Ovs82, Ibr85, Olv93, Olv95}. 
Researchers in these fields generally deal with the following three classification problems.
\begin{enumerate}
\item Classification of differential equations with Lie symmetry groups acting on a given space of dependent and independent variables.
\item Group classification of differential equations belonging to a specific class (the so-called \textit{preliminary group classification}).
\item Classification of differential equations with a prescribed abstract Lie symmetry group or  algebra.
\end{enumerate}

The main concept of solving the \textit{first classification problem} was formulated by Sophus Lie himself in his classic paper on the classification of ordinary differential equations admitting non-trivial symmetry groups~\cite{Lie1888}.
Recall that the first step of solving this problem consists of listing all the inequivalent Lie group actions on a given space of dependent and independent variables.
Following this, the second step is to calculate the differential invariants for each inequivalent Lie group action and then use them for constructing the classes of invariant differential equations. Following this procedure, Sophus Lie exhaustively listed all the inequivalent local Lie transformation groups of the complex plain and calculated all the corresponding differential invariants~\cite{Lie1888} (see also~\cite{Olv94} for a more modern viewpoint). 
Subsequently, Gonz\'alez-L\'opez \textit{et al.} obtained the analogous classification for local Lie group actions on $\mathbb{R}^2$~\cite{GonKamOlv92}, using which Nesterenko constructed bases of the corresponding differential invariants and operators of invariant differentiation~\cite{Nes06}. It also was Sophus Lie who successfully provided one of the first general solutions to the first classification problem applied to partial differential equations.
In his study~\cite{Lie81}, he performed a group classification of linear second-order partial differential equations in two independent variables.
It is important to note that solving the first classification problem for the above cases was largely possible owing to the small dimensions of the spaces of independent and dependent variables as the computational difficulties of applying the infinitesimal technique grow rapidly with increasing these dimensions.
 
Further systematic development of Lie's concepts was conducted by Ovsyannikov~\cite{Ovs59, Ovs82}. 
His approach, based on the use of \textit{equivalence transformations} for differential equations and the concept of the so-called \textit{arbitrary element}, made it possible to implement a large-scale program for solving the second classification problem applied to fundamental equations of mathematical physics.
In particular, Ovsyannikov himself has performed the group classification of a class of non-linear heat conductivity equations~\cite{Ovs59}, and subsequently Ibraginov \textit{et al}.~\cite{AkhGazIbr87} employed his approach for the classification of non-linear filtration equations.
The concepts of Ovsyannikov have also formed the basis of the \textit{preliminary group classification method}, which was first introduced in \cite{AkhGazIbr91} and was made well-known by Ibragimov \textit{et al}.~\cite{IbrTorVal91}.
Ibragimov and Torrisi applied this method to non-linear detonation equations~\cite{IbrTor92} and non-linear hyperbolic equations~\cite{IbrTorVal91}, and subsequently, Torrisi and coauthors obtained the preliminary group classification of some classes of diffusion and heat conduction equations~\cite{TorTraVal96, TorTra98}.
Furthermore, the preliminary group classification method was repeatedly modified, and the range of its applicability was constantly expanded (see, for example, \cite{BihCarPop12, VanBihPop20}).
In particular, we mention here a series of the papers~\cite{ZhdLah99, BasLahZhd01, LahZhd05, LahZhdMag06}, in which classification approaches based on a combination of Lie's infinitesimal method, Ovsyannikov's concept of the equivalence group, and the theory of Lie algebra realizations is suggested.

In practice, apart from the above classification problems, we often need to construct differential equations whose Lie symmetry groups are given abstractly, for example, by structure constants of the associated Lie algebras. 
Solving this classification problem plays a significant role in theoretical physics providing the basis for constructing physical models with prescribed symmetry groups.
The difficulty, however, is that we have previously to give the classification of all possible actions of a given Lie group $G$ in \textit{all} spaces with finite numbers of dependent and independent variables.
In the language of Lie algebras, this is equivalent to the problem of classification of all representations for the Lie algebra $\g = \mathrm{Lie}(G)$ by vector fields, which is quite complicated in general. 
If the classification of all the realizations is known, then we can list all the inequivalent differential equations with the symmetry Lie group $G$ by calculating the differential invariants for every inequivalent realization of the Lie algebra $\g$.  
This is also a quite difficult problem which is reduced to integrating systems of partial first-order differential equations.

Recently some new results have been obtained that can be used for developing an effective approach for the above-mentioned problems; this approach completely excludes the stages of integrating differential equations.

First, we would like to highlight the paper~\cite{PopBoyNesLut03}, in which Popovich and co-authors obtained the complete classification for local inequivalent realizations of real Lie algebras with dimensions no larger than four.
Secondly, significant progress has been also made in the theory of differential invariants.
Olver and Fels~\cite{FelOlv98, FelOlv99, Olv2007} proposed an effective geometric method to compute differential invariants for both finite-dimensional Lie group actions and infinite Lie pseudo-groups.
Their approach, based on works by Cartan and called the \textit{method of moving coframes}, allows the construction of differential invariants without a direct integration of differential equations.
As well as Lie's approach, the method of Olver and Fels is universal but unfortunately leads to cumbersome calculations even for low-dimensional groups. Although the use of symbolic calculation packages can simplify the problem (see, for example, Boyko \textit{et al}~\cite{BoyPatPop06}). 
Moreover, the application of this method requires knowledge of how a symmetry group acts on a space of dependent and independent variables in the explicit form.

A slightly different approach to the computation of differential invariants was proposed in~\cite{Shi07}.
In this study, the author considered the case when there is one invariant dependent variable $u$ and a transformation group $G$ acts transitively on the space $M$ of $n$ independent variables $\{x\}$.
The basic result, in this case, is that differential invariants and operators of invariant differentiation may be described in the general free-coordinate form.

The present paper is devoted to a further generalization of the ideas and techniques from \cite{Shi07}.
Our main observation is that the general form of differential invariants significantly depends on the type of group action, and therefore we begin our paper with an original classification of differential equations based on the regular actions of their symmetry groups.
In accordance with this classification, all partial differential equations are divided into five classes (types I -- V).
Further, we explore in detail the case when a symmetry group act freely and regularly (types I and II) and prove theorems described the general form of the corresponding differential invariants.
Based on these theorems and well-known classifications low-dimensional Lie algebras, we give a complete classification of all scalar second-order partial differential equations of types I and II with symmetry groups of dimensions $< 4$ (up to a ``hodograph'' transformation). 

The article is organized as follows.
In Section~\ref{sec:1}, we briefly recall the basic facts concerning differential equations and their symmetry groups in regard to differential geometry. In this section, we also suggest a classification of differential equations based on the types of symmetry group actions; this classification enables dividing differential equations in an invariant manner into five distinct types.
In the second section, we present Theorems \ref{t3} and \ref{t5} that provide an effective algorithm for the solution of the group classification problem of differential equations with regular symmetry groups acting intransitively and freely.
Based on these results, we solve the problem of the group classification of scalar second-order differential equations whose free intransitive symmetry groups have dimensions up to and including three.
In the third section, we introduce the concept of the covariant form of a differential equation and show that any scalar differential equation can be written in such a form.
We also formulate the necessary and sufficient condition under which a given equation is the covariant form of some differential equation.
Applying this algorithm, we present an exhaustive classification of both second-order differential invariants and quasi-linear second-order differential equations for all inequivalent two- and three-dimensional simply transitive symmetry groups, in Section~\ref{sec:4}.

\section{Problem statement}
\label{sec:1}

In this section, we recall the standard mathematical constructions and formulate the problems discussed in this study.
For more details of the group analysis of differential equations, see classical literature
~\cite{Ovs82, Ibr85, Olv93, Olv95}. 
The general information on Lie groups and their homogeneous spaces can be found, for example, in~\cite{War13, Che18, Hel79}.

Let $Z\subset \mathbb{R}^n$ be an open subset (regarded as the space of independent and dependent variables for some differential equation), and let $\Diff Z$ be the infinite-dimensional diffeomorphism group of~$Z$. 
The subset $Z$ looks locally like the Cartesian product $Z\simeq U\times X$, where $X$ and $U$ are the spaces of independent and dependent variables, respectively.
The space~$U$ can be uniquely extended to the $k$-jet space $U\to U_k=U\times\Uk{U}{1}\times \cdots \times\Uk{U}{k}$ with local coordinates $(u^\alpha,u_i^\alpha,u^\alpha_{ij},\dots,u^\alpha_{i_1\dots i_k})$.
A \textit{differential equation} (or a \textit{system of differential equations}) is a surface $\cal E$ in the extended space $Z_k\simeq U_k\times X$:
\begin{equation}\label{e1}
{\cal E}: \quad E_\mu(x,u,u_i,u_{ij},\dots)=0, \quad \mu=1,\dots,s. 
\end{equation}

Let $r_\mu$ be the maximal order of derivatives on which the $\mu$-th equation in \eqref{e1} depends. 
It is understandable that this number does not change under diffeomorphisms of $Z$ and the choice of the ``splitting'' $Z\simeq U\times X$. 
Thus, the $s$-tuple $\ord\cE= \{r_1,\dots,r_s\}$ is a $\Diff Z$--invariant characteristic of the equation~\eqref{e1}.

A \textit{solution} of the differential equation~\eqref{e1} is a surface $S$ in $Z$:
\[
S:\quad u^\alpha=f^\alpha(x),\quad \alpha=1,\dots \dim U.
\]
The codimension of this surface, $\codim S$, coincides with the number of dependent variables, and is also a $\Diff Z$--invariant characteristic of the differential equation~\eqref{e1}.
The extended surface $S_k\subset Z_k$,
\[
S_k:\quad
u_i^\alpha=\frac{\p f^\alpha(x)}{\p x^i },\quad
u_{ij}^\alpha=\frac{\p^2 f^\alpha(x)}{\p x^i\p x^j },\quad
\dots\ ;\quad (x,u)\in S,
\]
which could be regarded as an integral submanifold of the  distribution of Pfaffian forms
\[
du^\alpha-u_i^\alpha d x^i=0,\quad d u^\alpha_{i}-u^\alpha_{ij} d x^j=0,\quad
\dots,
\]
is a submanifold of the manifold $\cal E$.

Denote by $\mathrm{Sol}\,\cE$ the set of all solutions of the differential equation \eqref{e1}. 
By definition, the \textit{symmetry group} of this equation is a subgroup $G = \mathrm{Sym}\,\cE \subset \Diff Z$ that transforms the solutions of $\cE$ to other solutions, i.e. if $g\in \mathrm{Sym}\, \cE$ and $S \in \mathrm{Sol}\, \cE$, then we have $g(S)\in \mathrm{Sol}\,\cE$.
In this study, we restrict our attention to connected local Lie symmetry groups $G$, which are generated by the corresponding Lie algebras $\g$ of vector fields $X_i$ on $Z$:
\begin{equation}\label{e2}
X_i=\zeta^a_i(z)\,\frac{\p}{\p z^a}\in T_zZ;\quad [X_i,X_j]=C_{ij}^kX_k.
\end{equation}
Here, $C_{ij}^k$ are the structure constants of the Lie algebra~$\g$.

\begin{remark}
In the entire study, we assumed that the symmetry group $G$ acts \textit{regularly} on $Z$. 
Moreover, we also assume that the group act \textit{effectively}; otherwise, we can replace $G$ by the effectively acting quotient group $G/G_0$, where $G_0$ denotes the global isotropy subgroup of $Z$: $G_0 z = z$ for all $z \in Z$.
\end{remark}

Let $G$ be the symmetry group of a given differential equation $\cE$ and $X_i$ its infinitesimal generators forming the corresponding Lie algebra~$\g$.
Let $M=Gz_0$ be the orbit through a point $z_0\in Z$. 
It is known that there is a $G$-equivariant diffeomorphism between the homogeneous space $M$ and the set of right (or, equivalently, left) cosets $G/H$, where $H=H_{z_0}$ is the isotropy subgroup at the point~$z_0$. 
Owing to the regularity of the group action, we can locally introduce coordinates charts of the form $X\times Y$ in $Z$, where $x\in X$ are coordinates on the orbit $M$, and $y\in Y$ are additional coordinates in $Z$, which can be interpreted as invariants of the group action of~$G$ (see, for example,~\cite{Olv95}).
(It is important to distinguish between these coordinate charts and charts $Z\simeq U\times X$ obtained as a result of the ``splitting'' into dependent and independent variables.)
For brevity, we shall call $\{y\}$ the \textit{invariant coordinates}.

\begin{wrapfigure}{l}{0.45\textwidth}
\begin{center}
\includegraphics[width=0.35\textwidth]{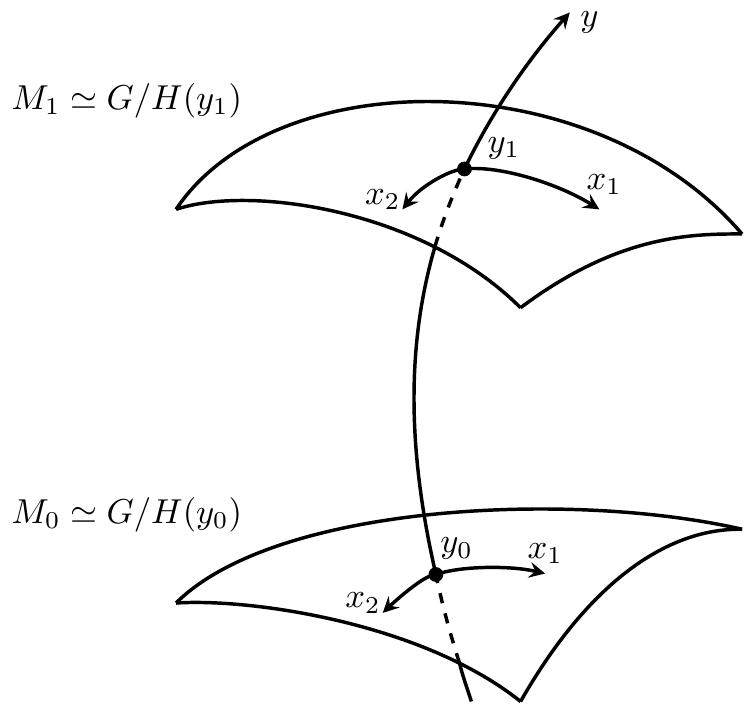}
\end{center}
\caption{\small Orbits of a group action can depend on the invariant variables~$y$}
\label{pic:1}
\end{wrapfigure}

Note that the isotropy subgroup $H$ and its Lie algebra $\h$, in general, depend on the invariant variables $y\in Y$, i.e. $H=H(y),\ \h=\h(y)$. 
From the geometric point of view, this means that when a point $z_0 \in Z$ moves to another point $z_1$ along a line transverse to the tangent space $T_{z_0} M_0$, the homogeneous space $M_0 = G z_0$  is mapped into the homogeneous space $M_1 = G z_1$ homeomorphic to $M_0$. 
However, algebraic properties of the latter can differ from those of the space $M_0$ owing to the inequation $\h(y_0)\neq \h(y_1)$ (see Fig.~\ref{pic:1}).
This, in turn, leads to the infinitesimal generators $X_i$ whose coefficients depend on the invariant variables  $\{ y \}$ in the coordinate chart $Z \simeq X\times Y$,
\[
X_i=\zeta^a_i(x,y)\frac{\p}{\p x^a}.
\]
Moreover, this dependence cannot be eliminated by coordinate transformations.
In this regard, we introduce a numerical characteristic $\ind (X,Y)$, called the \textit{coupling index}, whose definition will be provided in our subsequent papers.

In view of the above considerations, the (regular) actions of a connected Lie group $G$ can be divided into several types.
\begin{description}

\item[Type I.] \emph{Free action of $G$ on $Z$}.

In this case, $\rank\zeta_i^a(x)=\dim\g<\dim Z$, $\h=\{0\}$, and there are invariant variables $\{y\}$; however, coefficients $\zeta_i^a(x)$ do not depend on their ($\ind(X,Y)=0$).

\item[Type II.] \emph{Simply transitive action of $G$ on $Z$}.

In this case, $\rank\zeta_i^a(x)=\dim\g=\dim Z$, $\h=\{0\}$, and there are no invariant variables $\{y\}$.

\item[Type III.] \emph{Intransitive action of $G$ on $Z$ with a non-maximal coupling index}.

In this case, $\rank\zeta_i^a(x,y)=\dim\g/\h<\dim Z$, $\dim\h>0,\ \h=\h(y)$, and there are the invariant variables $\{y\}$, $\ind(X,Y)<\dim Y$.

\item[Type IV.]\emph{Transitive action of $G$ on $Z$}.

In this case, $\rank\zeta_i^a(x)=\dim\g/\h=\dim Z$, $\dim\h>0$, and there are no invariant variables $\{y\}$.

\item[Type V.] \emph{Intransitive action of $G$ on $Z$ with the maximal coupling index}.

In this case, $\rank\zeta_i^a(x,y)=\dim\g/\h<\dim Z$, $\dim\h>0,\ \h=\h(y)$, and there are the invariant variables $\{y\}$, $\ind(X,Y)=\dim Y$.

\end{description}

The usefulness of this classification is because the most effective methods for constructing invariant differential equations are essentially different for different types of group actions.

Thus, it is convenient to introduce the $\Diff Z$--invariant classification of differential equations given in Table~\ref{tab:1}.
In the table, symbols $n,m,l,k,r_j,s$, and $t$ are arbitrary positive integers.

\begin{table}[!h]
\small
\begin{center}
\label{tab:1}
\caption{\small Types of differential equations.}
\vspace{0.25cm}
\begin{tabular}{|c|c|c|c|c|c|c|}
\hline 
Type & $\dim Z$ & $\ord \cE$& $\codim S$ & $\dim\g$ & $\dim\h$ & $\ind (X,Y)$\\ 
\hline  
I & $n+l$ &  $\{r_1,\dots,r_t\}$ &$s$ & $n$ & 0 & 0 \\ 
\hline
II & $n$ &  $\{r_1,\dots,r_t\}$ &$s$ & $n$ & 0 & 0 \\ 
\hline
III & $n+l+k$ &  $\{r_1,\dots,r_t\}$ &$s$ & $n+m$ & $m$ &$k$  \\ 
\hline
IV & $n$ &  $\{r_1,\dots,r_t\}$ &$s$ & $n+m$ & $m$ & 0 \\ 
\hline
V & $n+k$ &  $\{r_1,\dots,r_t\}$ &$s$ & $n+m$ & $m$ &$k$  \\ 
\hline
\end{tabular} 
\end{center}
\end{table}

It is clear that the first position in Table~\ref{tab:1}, which indicates types of differential equations, is not informative and will frequently be omitted.
Also note that instead of specifying the numbers $\dim \g$ and $\dim \h$ one may describe the structures of Lie algebras $\g$ and their subalgebras $\h$ in more detail. 
For instance, this may be realized by specifying the commutation relations $[e_i,e_j]=C_{ij}^k e_k$ of a Lie algebra $\g$ in some basis $\< e_i \>$ and by the explicit listing the basis elements of its subalgebra: $\h=\< a_\alpha^i(y)e_i\> $.
It will be shown in our subsequent studies that this information is sufficient for the complete description of the class of differential equations of a given type.

\begin{remark}
The interested reader may find a discussion of different approaches to $\Diff Z$-invariant classification of differential equations in Ovsyannikov's book~\cite{Ovs82}.
In particular, Ovsyannikov himself suggested the classification to be conducted as follows. 
To each system of differential equations, we associate a quadruple of numbers $(\nu, \mu, \kappa, \sigma)$, where $\nu$ and $\mu$ are the numbers of independent and dependent variables, respectively,  
$\kappa$ is the maximal order of the derivatives that appear in the system, and $\sigma$ is the number of equations. 
Remarkably, the classification scheme outlined in Table~\ref{tab:1} is a more subtle systematization of differential equations than Ovsyannikov's one because it take into account the structures of symmetry group actions. 
The relation between Ovsyannikov's classification and our one is given as follows:
$$
\nu + \mu = \dim Z,\quad \mu=\codim S, \quad
\kappa = \max \ord \cE,\quad
\sigma =\dim \ord \cE=t.
$$
\end{remark}

Let us now formulate the basic problems of interest in this paper. 
Our principal goal is to provide a complete description of all non-linear partial differential equations whose types are presented in Table~\ref{tab:1}. 
In this study, we focus on the second-order scalar partial differential equations of types I ($\{n+m,\{2\},1,n,0,0\}$) and II ($\{n,\{2\},1,$ $n,0,0\})$. 

We note that the case of one invariant variable, regarded as a dependent one, was considered in~\cite{Shi07}.
According to our classification, the mentioned paper is devoted to types $\{n+1,\{2\},1,n,0,0\}$ (a special case of type I) and $\{n+1,\{2\},1,n+m,m,0\}$ (a special case of type III). 
Let us show that there is a fundamental difference between the equations of types $\{n,\{2\},1,n,0,0\}$ and $\{n+1,\{2\},1,n,0,0\}$. 

Let us consider the case of a scalar second-order differential equation with $n+1$ dependent and independent variables. 
In this case, an $n$-dimensional symmetry group $G$ acts on an $n$-dimensional space simply transitively ($\dim\h=0$), and there is one invariant variable $y_1$, which is declared, for convenience, as a dependent one, $u=y_1$. 
As the dimension of the $G$-orbits coincides with the dimension of $G$, we can assume that the group acts on itself by \textit{right multiplications} and that the infinitesimal generators of this action coincide with the \textit{left-invariant vector fields} on $G$: 
\begin{equation}\label{e3}
X_i=\xi_i=\sum\limits_{a=1}^{n}\xi_i^a(x)\frac{\p}{\p {x^a}},\quad i=1,\dots,n=\dim\g.
\end{equation}
(Here and elsewhere, symbol $\xi_i$ denotes the left-invariant vector field on~$G$ associated with the basis element $e_i \in \g$, $\xi_i(x) = (L_x)_* e_i$, $x \in G$).

Consider now an equation of type $\{n,\{2\},1,n,0,0\}$, i.e., the case of a scalar second-order differential equation with $n$ dependent and independent variables. 
In this case, an $n$-dimensional symmetry group $G$ acts simply transitively on an $n$--dimensional space $Z$ ($\dim\h=0$) and there are no invariant variables~$\{y\}$. 
Again, owing to the dimension of the orbits coinciding with the dimension of the group, it can be assumed that $G$ acts on itself by right multiplications and the infinitesimal generators \eqref{e2} of the action coincide with the left-invariant vector fields: $X_i=\xi_i=\xi_i^j(z)\p_{z^j}$. 
Since all variables $z^j\  (j=1,\dots,n)$ are treated equally, we can select the variable $z^n$ as an independent one and rewrite the generators $X_i$ in a relatively traditional form:
\begin{equation}\label{e4}
X_i=\sum\limits_{a=1}^{n-1}\xi^a_i(x,u)\frac{\p}{\p x^a}+\theta_i(x,u)\frac{\p}{\p u}, \quad i=1,\dots,n=\dim\g.
\end{equation}
Here, $\theta_i(x,u)=\xi^n_i(z),\ \xi^a_i(x,u)=\xi^a_i(z),\ a=1,\dots,n-1$.

Comparing Eqs.~\eqref{e3} and \eqref{e4}, it can be seen that the case of $G$-action with the generators~\eqref{e3} considered in \cite{Shi07} is much easier than the case of $G$-action with the generators~\eqref{e4}. 

It will be shown in Section~\ref{sec:2} that an arbitrary differential equation can be written in the so-called \textit{covariant form} preserved the original ``equivalence'' of the dependent and independent coordinates. 
This will allow us to describe $G$-invariant differential equations with the generators \eqref{e3} and \eqref{e4} in a unified manner.

\section{Differential equations with free symmetry groups}
\label{sec:5}

Suppose that a Lie group $G$ acts on a space $Z$ freely and intransitively (the case of free and transitive action will be considered below).
Locally, we have $Z\simeq X\times Y$, where $X$ is an orbit of the group action, and $Y$ is a subspace transverse to the orbit; coordinates on $Y$ are invariants of the action. 
Since $X$ is diffeomorphic to $G$, the infinitesimal generators of the group action have the form \eqref{e3} and coincide with the left-invariant vector fields $\xi_i$ on $G$. 
In contrast to the scenario examined in the study in \cite{Shi07}, here we consider a more general case when there are several invariant variables $\{y^1,\dots,y^m\}$.

The second-order differential equations with such symmetry groups have type I: $\{n+m,\{2\},1,n,0,0\}$. 
Without the loss of generality, we can regard the invariant variable $y^m$ as the dependent variable~$u$.
Indeed, if the dependent variable is not invariant (for example, $u=x^n$), a hodograph transformation can then be used to make the invariant variable $y^m$ dependent.
(We note that hodograph transformations have the property of mapping quasi-linear differential equations to quasi-linear ones). 

Let us realize the Lie algebra $\g = \mathrm{Lie}(G)$ by the right-invariant vector fields $\eta_i(z) = (R_{z^{-1}})_* e_i = \eta_i^j(z)\p_{z^j}$.
It is well-known that left- and right-invariant vector fields on $G$ satisfy the commutation relations
\begin{equation*}
[\xi_i, \xi_j] = C_{ij}^k \xi_k,\quad
[\eta_i, \eta_j] = - C_{ij}^k \eta_k,\quad
[\xi_i, \eta_j] = 0.
\end{equation*}
We will also use the notation
\begin{equation}\label{ed1}
\uu{i}=\hat{\eta}_i u=\eta_i^j(x)u_j,\quad \uu{ij}=\frac12(\hat{\eta}_i\hat{\eta}_j+\hat{\eta}_j\hat{\eta}_i)u=(\hat{\eta}_i\circ\hat{\eta}_j)u,\quad
\uu{j_1\dots j_k}=\frac{1}{k!}\sum\limits_{\sigma\in S_k} \hat{\eta}_{j_{\sigma_1}}\cdots \hat{\eta}_{j_{\sigma_k}}u,
\end{equation}
where $\hat{\eta}_i=\eta_i^j D_j$ and $D_j=\p_{z^j}+u_j\p_u+u_{jk}\p_{u_k}+\cdots$ is the total derivative operator.

The following theorem was proved in \cite{Shi07}.

\begin{theorem}
\label{t3}
Let 
\begin{equation}
\label{magazev:1}
E(x,u,\Uk{u}{1},\Uk{u}{2},\dots,\Uk{u}{k})=0
\end{equation}
be a differential equation with $n$ independent variables $\{x^i\}$ and one dependent variable $\{u\}$ admitting an $n$-dimensional Lie symmetry group $G$ with the generators $X_i=\xi_i=\xi_i^j(x)\p_{x^j}$. Assume that $G$ acts simply transitively and effectively on the space of independent variables $x$, i.e. $\det \| \xi_i^j(x) \|\neq 0$. Then the differential equation \eqref{magazev:1} can be written in the form
\begin{equation}\label{e24}
F(u,\uu{i},\uu{ij},\dots,\uu{j_1\dots j_k})=0.
\end{equation}
In other words, the collection of functions $u,\uu{i},\uu{ij},\dots,\uu{j_1\dots j_k},\dots$ forms a basis of the differential invariants for the $G$-action and $\hat{\eta}_i$ are the corresponding operators of invariant differentiation.
\end{theorem}

This theorem can be easily generalized to the case where the number $m$ of invariant variables is arbitrary. As we have noted, we assume that $y^m = u$ and therefore the rest invariant variables $y^1, \dots, y^{m-1}$ are regarded as independent ones.
It is easy to see that the total derivative operators $D_{y^\mu}$ are operators of invariant differentiation. 
Thus, to the differential invariants $y^\mu$, $u$, $u_{(i)}$, $u_{(ij)}$, $\dots$, we have to add the invariants $u_{y^\mu}$, $u_{y^\mu y^\nu}$, $u_{(i)y^\mu}$ etc.

Hence we can formulate 

\begin{theorem}\label{t5}
Any scalar second-order partial differential equation whose symmetry group acts freely and intransitively can be written in the form:
\begin{equation*}
F(y,u,u_{y^\mu},\uu{i},u_{y^\mu y^\nu},\uu{ij},\uu{i}{}_{y^\mu})=0.
\end{equation*}
In particular, any scalar second-order partial differential equation linear in the highest derivatives is represented as follows:
\begin{equation}\label{e28}
 \sum\limits_{\mu,\nu=1}^{m-1}a^{\mu \nu}u_{y^\mu y^\nu}+
 \sum\limits_{i,j=1}^{n}b^{ij}\uu{ij}+
 \sum\limits_{\mu=1}^{m-1}\sum\limits_{i=1}^n c^{\mu i}\uu{i}{}_{y^\mu}+d=0.
 \end{equation} 
Here, $a^{\mu \nu},b^{ij}, c^{\mu i},d$ are arbitrary functions of $\{ y^\mu,u,u_{y^\mu},\uu{i} \}$, $\mu = 1, \dots, m-1$, $i = 1, \dots, n$.
\end{theorem}
 
Let us list all the inequivalent scalar second-order differential equations of this type for the Lie symmetry groups of dimensions no larger than three. 
In the tables below, we indicate the Lie algebras of the symmetry groups, their infinitesimal generators, and all  the second-order differential invariants.
The invariant quasi-linear differential equations have the same structure as~\eqref{e28}.
Everywhere below we use the notation: $y=\{y^1,\dots,y^{m-1}\}$, $u_{\mu}=u_{y^\mu},\; u_i=u_{x^i},\; u_{\mu\nu}=u_{y^\mu y^\nu},\; u_{i\mu }=u_{x^i y^\mu },\;u_{ij}=u_{x^ix^j}\; (i,j=1,\dots,n;\; \mu,\nu=1,\dots,m-1).$

\subsection{One-dimensional symmetry groups (type $\{1+m,\{2\},1,1,0,0\}$)}

Up to isomorphism, there is one one-dimensional Lie algebra $\g^1 \simeq \mathbb{R}$.

\begin{table}[!h]
\centering
\small
\caption{\small Second-order differential invariants for the one-dimensional Lie algebra $\g^1$.}\label{tab:4}
\vspace{4mm}
\begin{tabular}{|l|l|l|}
\hline 
Lie algebra &  Generators $X_i$ & Second-order differential invariants \\ \hline 
$\g^1$ &  $X_1=\p_{x^1}$ & $y,u,u_1,u_{\mu},u_{11},u_{1\mu},u_{\mu\nu}$ \\ 
\hline 
\end{tabular} 
\end{table}

\subsection{Two-dimensional symmetry groups (type $\{2+m,\{2\},1,2,0,0\}$)}

There are only two non-isomorphic two-dimensional Lie algebras: the abeliwn two-dimensional Lie algebra $2 \g^1 \simeq \mathbb{R}^2$ and the Lie algebra $\g^2$ of the affine group $\mathrm{Aff}(\mathbb{R})$.

\begin{table}[!h]
\centering
\small
\caption{\small Second-order differential invariants for two-dimensional Lie algebras.}\label{tab:5}
\vspace{4mm}
\begin{tabular}{|l|l|l|}
\hline 
Lie algebras &  Generators $X_i$ &	Second-order differential invariants \\ \hline 
$2\g^1$ &  $X_1=\p_{x^1},X_2=\p_{x^2}$ & $y,u,u_1,u_2,u_{\mu},u_{11},u_{12},u_{22},u_{1\mu},u_{2\mu},u_{\mu\nu}$ \\ 
\hline 
$\g^2$ 	&	$X_1=\p_{x^1}$, $X_2 = x^1 \p_{x^1} + \p_{x^2}$	&	$y$, $u$, $e^{x^2}u_1,$ $u_2,$ $u_{\mu},$ $e^{2x^2}u_{11}$, $e^{x^2} \left ( u_{12} + \frac{u_1}{2} \right )$, $u_{22}$, $e^{x^2}u_{1\mu}$, 	\\ 
$[e_1,e_2] = e_1$	&	&	 $u_{2\mu}$, $u_{\mu\nu}$	\\
\hline
\end{tabular} 
\end{table}

\subsection{Three-dimensional symmetry groups (type $\{3+m,\{2\},1,3,0,0\}$)}

Here we use the classification of three-dimensional Lie algebras suggested by Mu\-ba\-rak\-zya\-nov~\cite{Mub63}. In Table~\ref{tab:6}, we use the notation:
\begin{multline*}
u^7_{(11)} = u_{11}\,\frac{\cos^2 x^3}{\cos^2 x^2} - u_{12}\, \frac{\sin 2 x^3}{\cos x^2} + 2 u_{13}\, \frac{\tan x^2 \cos^2 x^3}{\cos x^2} + u_{22} \sin^2 x^3 - u_{23} \tan x^2 \sin 2 x^3 - 
\\
- u_1\,\frac{\tan x^2 \sin 2 x^3}{\cos x^2} - u_2 \tan x^2 \cos^2 x^3 - u_3 \left ( \frac{1}{2} + \tan^2 x^2 \right ) \sin 2 x^3,
\end{multline*}
\begin{multline*}
u^7_{(12)} = u_{11}\,\frac{\sin 2 x^3}{2 \cos^2 x^2} + u_{12}\, \frac{\cos 2 x^3}{\cos x^2} + u_{13}\, \frac{\tan x^2 \sin 2 x^3}{\cos x^2} - \frac{1}{2}\, u_{22} \sin 2 x^3 + u_{23} \tan x^2 \cos 2 x^3 + 
\\
+ u_1\,\frac{\tan x^2 \cos 2 x^3}{\cos x^2} - \frac{1}{2}\, u_2 \tan x^2 \sin 2 x^3 + u_3 \left ( \frac{1}{2} + \tan^2 x^2 \right ) \cos 2 x^3,
\end{multline*}
\begin{multline*}
u^7_{(22)} = u_{11}\,\frac{\sin^2 x^3}{\cos^2 x^2} + u_{12}\, \frac{\sin 2 x^3}{\cos x^2} + 2 u_{13}\, \frac{\tan x^2 \sin^2 x^3}{\cos x^2} + u_{22} \cos^2 x^3 + u_{23} \tan x^2 \sin 2 x^3 + 
\\
+ u_1\,\frac{\tan x^2 \sin 2 x^3}{\cos x^2} - u_2 \tan x^2 \sin^2 x^3 + u_3 \left ( \frac{1}{2} + \tan^2 x^2 \right ) \sin 2 x^3.
\end{multline*}

\begin{center}
\small
\centering
\LTcapwidth=\textwidth
\begin{longtable}{|l|l|l|}
\caption{\small Second-order differential invariants for three-dimensional Lie algebras.}\label{tab:6}\\
\hline
Lie algebra &  Generators $X_i$ & Second-order differential invariants \\
\hline
$3\g^1$	&	$X_1 = \p_{x^1}$, $X_2 = \p_{x^2}$, & $y$, $u$,	$u_1$, $u_2$, $u_3$, $u_\mu$, $u_{11}$, $u_{12}$, $u_{13}$, $u_{22}$, $u_{23}$,	\\
		&	$X_3 = \p_{x^3}$	&	$u_{33}$, $u_{1\mu}$, $u_{2\mu}$, $u_{3\mu}$, $u_{\mu\nu}$	\\
\hline
$\g^1+\g^2$	&	$X_1=\p_{x^1}$, &	$y$, $u$,	$e^{x^2}u_1$, $u_2$, $u_3$, $u_\mu$,  \\
$[e_1,e_2]=e_1$	&	$X_2=x^1 \p_{x^1}+\p_{x^2}$,	&	$e^{2x^2}u_{11}$, $e^{x^2}\left ( u_{12} + \frac{u_1}{2}\right )$, $e^{x^2}u_{13}$, $u_{22}$, $u_{23}$, $u_{33}$, 	\\
	&	$X_3=\p_{x^3}$	&	$e^{x^2}u_{1\mu}$, $u_{2\mu}$, $u_{3\mu}$, $u_{\mu\nu}$	\\
\hline
$\g^3_1$ 	&	$X_1=\p_{x^1}$, & $y$, $u$,	$u_1$, $x^3 u_1 + u_2$, $u_3$, $u_\mu$, $u_{11}$, $x^3 u_{11}+u_{12}$,  \\
$[e_2,e_3]=e_1$	&	$X_2=\p_{x^2}$,	& $u_{13}$, $(x^3)^2 u_{11} + 2 x^3 u_{12}+u_{22}$, $x^3 u_{13}+u_{23}+\frac{u_1}{2}$, 	\\
	&	$X_3=x^2\p_{x^1}+\p_{x^3}$	&	$u_{33}$, $u_{1\mu}$, $x^3 u_{1\mu}+u_{2\mu}$, $u_{3\mu}$, $u_{\mu\nu}$\\
\hline
$\g^3_2$ &	$X_1=\p_{x^1}$, & $y$, $u$,	$e^{x^3}u_1$, $e^{x^3}\left ( x^3 u_1 + u_2\right )$, $u_3$, $u_\mu$,  \\
$[e_1,e_3]=e_1$ &	$X_2=\p_{x^2}$, & $e^{2x^3}u_{11}$, $e^{2x^3}\left ( x^3 u_{11}+u_{12} \right )$, $e^{x^3}\left ( u_{13} + \frac{u_1}{2}\right )$,  	\\
$[e_2,e_3]=e_1+e_2$	&	$X_3=(x^1+x^2)\p_{x^1}+\hfill$	&	$u_{33}$, $e^{x^3}u_{1\mu}$, $e^{2x^3}\left ( (x^3)^2 u_{11} + 2 x^3 u_{12} + u_{22} \right )$, 	\\
	&	$\hfill+x^2\p_{x^2}+\p_{x^3}$	& $e^{x^3} x^3\left ( u_{13} + \frac{u_1}{2} \right ) + e^{x^3}\left ( u_{23} + \frac{u_1+u_2}{2} \right )$, 	\\
	&	&	$e^{x^3} \left ( x^3 u_{1\mu}+u_{2\mu}\right )$, $u_{3\mu}$, $u_{\mu\nu}$	\\
\hline
$\g^3_3$ &	$X_1=\p_{x^1}$, & $y$, $u$,	$e^{x^3}u_1$, $e^{x^3}u_2$, $u_3$, $u_\mu$, $e^{2x^3}u_{11}$, $e^{2x^3}u_{12}$,  \\
$[e_1,e_3]=e_1$	&	$X_2=\p_{x^2}$, & $e^{x^3} \left ( u_{13} + \frac{u_1}{2} \right )$, $e^{2 x^3}u_{22}$, $e^{x^3} \left ( u_{23} + \frac{u_2}{2} \right )$, $u_{33}$, 	\\
$[e_2,e_3]=e_2$	&	$X_3=x^1\p_{x^1}+x^2\p_{x^2}+\p_{x^3}$	& $e^{x^3}u_{1\mu}$, $e^{x^3}u_{2\mu}$, $u_{3\mu}$, $u_{\mu\nu}$	\\
\hline
$\g^3_4$ &	$X_1=\p_{x^1}$, & $y$, $u$, $e^{x^3}u_1$, $e^{h x^3}u_2$, $u_3$, $u_\mu$, $e^{2x^3}u_{11}$,  \\
$[e_1,e_3]=e_1$ &	$X_2=\p_{x^2}$, & $e^{(1+h)x^3}u_{12}$, $e^{x^3} \left ( u_{13} + \frac{u_1}{2} \right )$, $e^{2 h x^3}u_{22}$,  	\\
$[e_2,e_3]=he_2$ &	$X_3=x^1\p_{x^1}+\hfill$	&	$e^{h x^3} \left ( u_{23} + \frac{h u_2}{2} \right )$, $u_{33}$, $e^{x^3}u_{1\mu}$, $e^{h x^3}u_{2\mu}$, 	\\
$|h|\leq 1,$ $h \neq 0,1$	&	$\hfill+h x^2\p_{x^2}+\p_{x^3}$	&	$u_{3\mu}$, $u_{\mu\nu}$	\\
\hline
$\g^3_5$ &	$X_1 = \p_{x^1}$, &	$y$, $u$, $u_3$, $u_\mu$, $e^{2px^3}\left ( u_{11} + u_{22} \right )$,  \\
$[e_1,e_3]=pe_1-e_2$	&	$X_2 = \p_{x^2}$, 	&	$e^{px^3}\left ( u_1 \cos x^3 - u_2 \sin x^3 \right )$,	\\
$[e_2,e_3]=pe_2+e_1$ 	&	$X_3 = (p x^1 + x^2)\p_{x^1}+\hfill$	&	 $e^{px^3}\left ( u_1 \sin x^3 + u_2 \cos x^3 \right )$, 	\\
$p\geq 0$	&	$\hfill+(px^2-x^1)\p_{x^2}+\p_{x^3}$	&	$e^{2px^3} \left [ ( u_{11} - u_{22} ) \cos 2 x^3 - 2 u_{12} \sin 2 x^3 \right ]$, 	\\ 
	&	&	$e^{2px^3} \left [ ( u_{11} - u_{22} ) \sin 2 x^3 + 2 u_{12} \cos 2 x^3 \right ]$, 	\\
	&	&	$e^{px^3} \left [ \left ( u_{13} + \frac{pu_1-u_2}{2} \right ) \cos x^3 - \left ( u_{23} + \frac{pu_2+u_1}{2} \right ) \sin x^3 \right ]$,  	\\
	&	&	$e^{px^3}  \left [ \left ( u_{13} + \frac{pu_1-u_2}{2} \right ) \sin x^3 + \left ( u_{23} + \frac{pu_2+u_1}{2} \right ) \cos x^3 \right ]$, \\
	&	&	$e^{px^3}\left ( u_{1\mu} \cos x^3 - u_{2\mu} \sin x^3 \right )$, \\
	&	&	$e^{px^3}\left ( u_{1\mu} \sin x^3 + u_{2\mu} \cos x^3 \right )$, $u_{33}$, $u_{3\mu}$, $u_{\mu\nu}$\\
\hline
$\g^3_6$ 	&	$X_1=\p_{x^1},$ &	$y$, $u$, $e^{x^2}u_1 + 2 x^3 u_2 + (x^3)^2 u_3$, $u_2 + x^3 u_3$, $u_3$, $u_\mu$,  \\
$[e_1,e_2]=e_1$ &	$X_2=x^1\p_{x^1}+\p_{x^2}$, 	&	$ \frac{1}{4}\,e^{2x^2} u_{11} + e^{x^2} x^3 \left ( u_{12} + \frac{x^3 u_{13} + u_1}{2} \right ) + \hfill$ 	\\
$[e_1,e_3]=2e_2$	&	$X_3 = (x^1)^2 \p_{x^1} + \hfill$	& $\hfill+(x^3)^2 \left ( x^3 u_{23} + u_{22} + \frac{u_2 + x^3 u_3}{2} + \frac{(x^3)^2 u_{33}}{4} \right )$,	\\
 $[e_2,e_3]=e_3$	&	$\hfill+2 x^1 \p_{x^2} +e^{x^2} \p_{x^3}$	&	$e^{x^2} \left ( u_{12} + x^3 u_{13} + \frac{u_1}{2} \right ) + (x^3)^3 u_{33} + \hfill$ 	\\
 	&	&	$\hfill+3 (x^3)^2 \left ( u_{23} + \frac{u_3}{2} \right ) + x^3 ( 2 u_{22} + u_2)$,	\\
 	&	&	$e^{x^2} u_{13} + (x^3)^2 u_{33}+x^3(2u_{23}+u_3)+u_2$,\\
 	&	&	$(x^3)^2u_{33}+x^3(2u_{23}+u_3)+u_{22}$, \\
 	&	&	$x^3 u_{33}+u_{23}+\frac{u_3}{2}$, $u_{33}$, $x^3 u_{3\mu}+u_{2\mu}$,  \\
 	&	&	$e^{x^2}u_{1\mu}+2x^3u_{2\mu} + (x^3)^2 u_{3\mu}$, $u_{3\mu}$, $u_{\mu\nu}$\\
\hline
$\g^3_7$ &	$X_1 = \p_{x^1}$, &	$y$, $u$, $u_\mu$, $u_3$,	\\
$[e_1,e_2]=e_3$		&	$X_2=\sin x^1 \tan x^2 \p_{x^1} + \hfill$ 	&	$u_1\, \frac{\cos x^3}{\cos x^2} - u_2 \sin x^3 + u_3 \tan x^2 \cos x^3$,	\\
$[e_2,e_3]=e_1$ 	&	$\hfill+\cos x^1 \p_{x^2} + \frac{\sin x^1}{\cos x^2}\, \p_{x^3}$,	&	$u_1\, \frac{\sin x^3}{\cos x^2} + u_2 \cos x^3 + u_3 \tan x^2 \sin x^3$,	\\
$[e_3,e_1]=e_2$		&	$X_3=\cos x^1 \tan x^2 \p_{x^1} - \hfill$	&	$u^7_{(11)}$, $u^7_{(12)}$, $u^7_{(22)}$, $u_{33}$, $u_{3\mu}$, $u_{\mu\nu}$,		\\
					&	$\hfill-\sin x^1 \p_{x^2} + \frac{\cos x^1}{\cos x^2}\, \p_{x^3}$	&	$u_{13}\,\frac{\cos x^3}{\cos x^2} - u_{23}\sin x^3 + u_{33}\cos x^3 \tan x^2 - \hfill$	\\
					&	&	$\hfill - \frac{u_1}{2}\, \frac{\sin x^3}{\cos x^2} - \frac{u_2}{2}\, \cos x^3 - \frac{u_3}{2}\, \sin x^3 \tan x^2 $,	\\
					&	&	$u_{13}\,\frac{\sin x^3}{\cos x^2} + u_{23} \cos x^3 + u_{33} \sin x^3 \tan x^2 + \hfill$ \\
					&	&	$\hfill + \frac{u_1}{2}\, \frac{\cos x^3}{\cos x^2} - \frac{u_2}{2}\, \sin x^3 + \frac{u_3}{2} \cos x^3 \tan x^2 $,\\
					&	&	$u_{1\mu}\,\frac{\cos x^3}{\cos x^2} - u_{2\mu} \sin x^3 + u_{3\mu}\cos x^3 \tan x^2$,	\\
					&	&	$u_{1\mu}\,\frac{\sin x^3}{\cos x^2} + u_{2\mu} \cos x^3 + u_{3\mu}\sin x^3 \tan x^2$	\\
\hline
\end{longtable}
\end{center}

\section{Covariant form of differential equations}
\label{sec:2}

In this study, we explore the classes of differential equations that are orbits of the diffeomorphism group $\Diff Z$.
However, ``splitting'' the space $Z$ into dependent and independent variables is not a $\Diff Z$-invariant procedure.
To illustrate this, we consider the following example.

Let $(x,y)$ be local coordinates in a two-dimensional domain $Z \subset \mathbb{R}^2$, where $x$ is an independent variable and $y$ is a dependent one. 
Let $g\in \Diff Z$ be the diffeomorphism $(x,y) \mapsto (p,q) \equiv g(x,y)=(y+x/2,y-x)$. 
We note that in this case, we cannot answer the question: what is the image of a function $E=E(x,y,y_x,y_{xx})$ under the diffeomorphism $g \in \Diff Z$? 
Indeed, it may be a function $\tilde{E}(p,q,p_q,p_{qq})$ or function $\tilde{E}(p,q,q_p,q_{pp})$, or some other one. 
In fact, to answer this question, we need to again divide the coordinates on $Z$ into dependent and independent variables after acting the diffeomorphism.

It should be emphasized that if the diffeomorphism is an element of the one-parameter group generated by a vector field on $Z$, one generally accepts the rule that the division into dependent and independent variables is defined by continuity.
For instance, if the diffeomorphism $g \in \Diff Z$ is generated by the vector field
\[
X=\eta(x,y)\frac{\p}{\p y}+\zeta(x,y)\frac{\p}{\p x},\quad g_a(x,y)=(x_a,y_a),\  g=g_1;
\]
\[
\frac{\p y_a}{\p a}=\eta(x_a,y_a),\quad \frac{\p x_a}{\p a}=\zeta(x_a,y_a),\quad y_a|_{a=0}=y,\ x_a|_{a=0}=x,
\]
then, by continuity, we assume that $y_a$ are the dependent variables for all $a>0$.

Let us show that any differential equation can be rewritten in a special form (the covariant form) that preserves the original symmetry between the dependent and independent variables.

First, we formulate the main idea. 
Instead of the ``splitting'' $Z\simeq U\times X$, which is non-invariant under $\Diff Z$, we introduce the invariant ``splitting'' $Z\times W$, where the extra variables $w = (w^\alpha) \in W$ are invariant under the group $\Diff Z$. 
Our aim is to show that the differential equation~\eqref{e1} can be represented in the $\Diff Z$--invariant form
\begin{equation}\label{e5}
\tilde{\cE}: \quad \tilde{E}_\mu(z,w_i,w_{ij},\dots)=0,\quad \mu=1,\dots,s,
\end{equation}
where the functions $w^\alpha$ satisfy the additional conditions
\begin{equation}\label{e6}
w^\alpha=0, \quad \alpha=1,\dots,\dim W.
\end{equation}

\emph{Example 1.}\,$\triangleleft$
As a simple illustration of this idea, we consider a quasi-linear first-order partial differential equation of type $\{n,\{1\},1,*,*,*\}$:
\begin{equation}\label{e7}
\cE:\quad \sum\limits_{i=1}^{n-1}a^i(x,u)\frac{\p u}{\p x^i} + b(x,u)=0.
\end{equation}
The main trick of reducing \eqref{e7} to a linear partial differential equation is well-known; instead of the equation \eqref{e7}, one considers the equation
\begin{equation}\label{e8}
\tilde{\cE}:\quad \sum\limits_{i=1}^{n-1}a^i(z)\frac{\p w}{\p z^i} - b(z)\frac{\p w}{\p z^n}=0,
\end{equation}
where $z^n=u,\ z^i=x^i$ for $i=1,\dots,n-1$. 
If $w=w(z)$ is a solution of \eqref{e8}, then the equality $w(z)=0$ determines a solution of \eqref{e7}. 
We say that the differential equation \eqref{e8} (together with the condition $w=0$) is the \textit{covariant form} of the differential equation~\eqref{e7}. 
$\hfill\triangleright$

It is not difficult to generalize this trick for an arbitrary first-order partial differential equation.

\begin{proposition}
\label{pro:1}
A first-order partial differential equation
\begin{equation}\label{e9}
E(x,u;u_a)=0
\end{equation}
can be represented in the covariant form
\begin{eqnarray}\label{e10}
\tilde{E}(z;\Uk{w}{1})&=&0,\\ \label{e11}
w&=&0.
\end{eqnarray}
\end{proposition}

\emph{Proof.} $\triangleleft$ Assuming, by definition, that \eqref{e11} holds, we obtain
\[
dw=w_idz^i=w_adx^a+w_ndu=0.
\]
(Here and below we will use the notation: $n=\dim Z$, $u=z^n,\ x^a=z^a,\ a=1,\dots,n-1$). 
By virtue of $du-u_adx^a=0$, we have
\begin{equation}\label{e12}
u_a=-w_a/w_n.
\end{equation}
Thus, we have obtained the well-known rule of implicit differentiation. 
Substituting \eqref{e12} into \eqref{e9}, we deduce the equation
\[
\tilde{E}(z;\Uk{w}{1})=E(z;-w_a/w_n),
\]
which, together with the condition \eqref{e10}, is the covariant form of the first-order partial differential equation \eqref{e9}. $\hfill\triangleright$

The following example illustrates the content of Proposition~\ref{pro:1}.

\emph{Example 2.} $\triangleleft$
Let us consider a first-order partial differential equation, which is linear with respect to the  derivatives of $u$:
\[
g^{ab}(x,u)u_au_b+b(x,u)=0.
\]
Using \eqref{e12}, we can rewrite it as
\[
g^{ab}(z)w_aw_b+w_n^2b(z)=0.
\]
Together with the condition \eqref{e10}, this leads to the covariant form of the given first-order partial differential equation.
$\hfill\triangleright$

Recall that a function $f=f(x),\ x\in \R^n$ is said to be \textit{homogeneous of degree $k$}, if $f(\lambda x)=\lambda^k f(x)$ for all real numbers $\lambda$.

\begin{proposition}
Let $\tilde{E}(z;\Uk{w}{1})$ be a function that is homogeneous with respect to the derivatives~$w_i$.
The differential equation \eqref{e10} together with the condition \eqref{e11} is the covariant form of the differential equation \eqref{e9}, where $E(x,u;u_a)=\tilde{E}(x,u;-u_a,1)$.
\end{proposition}

\emph{Proof.} $\triangleleft$ 
Applying \eqref{e12} and using the homogeneity of $\tilde{E}$, we obtain
\[
\tilde{E}(z;w_i)=\tilde{E}(z;w_a,w_n)=\tilde{E}(z;-w_n u_a,w_n)=w_n^k\tilde{E}(x,u;-u_a,1).
\]
Thus, if $\tilde{E}(z;\Uk{w}{1})$ is a homogeneous function of the derivatives~$w_i$, then the  equation \eqref{e10} together with \eqref{e11} is the covariant form of the differential equation $\tilde{E}(x,u;-u_a,1) = 0$.$\hfill\triangleright$

\emph{Example 3.} $\triangleleft$ Let us consider a differential equation of the form \eqref{e10}, where $\tilde{E}(z,w_i)$ is a function that is homogeneous of degree two with respect to the second derivatives:
$$
g^{ij}(z)w_iw_j=0.
$$
By virtue of \eqref{e12}, we can write this in the form
$$
g^{na}(z)w_nw_a+g^{nn}(z)w_n^2 = w_n^2\left(g^{ab}(z)u_au_b-2g^{na}(z)u_a+g^{nn}(z)\right) = 0,
$$
which is equivalent to the differential equation
$$
g^{ab}(x,u)u_au_b-2g^{na}(x,u)u_a+g^{nn}(x,u)=0.
$$
$\hfill\triangleright$

Now, we consider second-order partial differential equations in detail.

\begin{proposition}\label{u3}
An arbitrary second-order partial differential equation
\begin{equation}\label{e13}
E(x,u;u_a;u_{ab})=0.
\end{equation}
can be represented in the covariant form
\begin{eqnarray}\label{e14}
\tilde{E}(z;\Uk{w}{1};\Uk{w}{2})&=&0,\\ \label{e15}
w&=&0.
\end{eqnarray}
\end{proposition}

\emph{Proof.} $\triangleleft$
Using the Pfaffian forms
\begin{eqnarray*}
dw-w_idz^i=0,\quad dw_i-w_{ij}dz^j=0,\quad i,j=1,\dots,n; \\
du-u_a dx^a=0,\quad du_a-u_{ab}dx^b=0,\quad a,b=1,\dots,n-1,
\end{eqnarray*}
we obtain the expressions for the second derivatives of the function $u = u(x)$ determined implicitly by the relation $w = 0$:
\begin{equation}\label{e16}
u_{ab}=-\frac{w_{ab}}{w_n}+\frac{w_{na}w_b}{w_n^2}+\frac{w_{nb}w_a}{w_n^2}-\frac{w_aw_bw_{nn}}{w_n^3}.
\end{equation}
Substituting \eqref{e12} and \eqref{e16} into \eqref{e13} leads to a differential equation of the form~\eqref{e14}. $\hfill\triangleright$

\begin{theorem}\label{t1}
The differential equation \eqref{e14} together with the condition \eqref{e15} is the covariant form of a second-order partial differential equation if and only if it is invariant under the infinite-dimensional group of the transformations:
\begin{equation}\label{e17}
w\to s(z)w.
\end{equation}
\end{theorem}

\emph{Proof.} $\triangleleft$ 
Let us assume that the differential equation \eqref{e14} together with the condition \eqref{e15} is the covariant form of a differential equation of the form~\eqref{e13}. 
In accordance with Proposition~\ref{u3}, \eqref{e14} is obtained by substituting \eqref{e12} and \eqref{e16} ($u_a\to u_a[w],\ u_{ab}\to u_{ab}[w]$) into \eqref{e13}. 
It is quite easy to check that the right-hand sides of the substitutions \eqref{e12} and \eqref{e16} are invariant under the transformations~\eqref{e17}: 
\[
u_a[s(z)w]=-\frac{D_{z^a}(s(z)w)}{D_{z^n}(s(z)w)}=-\frac{s_a(z)w+s(z)w_a}{s_n(z)w+s(z)w_n}=-\frac{w_a}{w_n}
=u_a[w].
\]
Here, we have used the condition \eqref{e15}. 
Analogously, it can be proved that equality $u_{ab}[s(z)w]=u_{ab}[w]$ holds.

Now, suppose \eqref{e14} and \eqref{e15} are invariant under the transformations \eqref{e17}.
Because the infinitesimal generator of these transformations has the form $X=\sigma(z)\,w\,\p_w$, the invariance of \eqref{e14} is equivalent to the equality
\[
\Uk{X}{2}\tilde{E}(z;\Uk{w}{1};\Uk{w}{2})|_{\tilde{E}=0} = \left(\sigma(z) D + \sum\limits_j\sigma_j(z) R_j\right)\tilde{E}(z;\Uk{w}{1};\Uk{w}{2})|_{\tilde{E}=0}=0,
\]
where we have used the notation
\[
D=\sum\limits_i w_i\frac{\p}{\p w_i}+\sum\limits_{i\leq j}w_{ij}\frac{\p}{\p w_{ij}},\quad
R_j=\sum\limits_i(1+\delta_{ij})w_i\frac{\p}{\p w_{ij}}.
\]
Thus, the invariance of \eqref{e14} under the transformations \eqref{e17} is equivalent to the equalities
\begin{eqnarray}\label{e18}
D\tilde{E}(z;\Uk{w}{1};\Uk{w}{2})&=&k \tilde{E}(z;\Uk{w}{1};\Uk{w}{2}),\\ \label{e19}
R_j \tilde{E}(z;\Uk{w}{1};\Uk{w}{2})&=&0,\quad j=1,\dots,n.
\end{eqnarray}

The equation \eqref{e18} is the requirement that the function $\tilde{E}(z;\Uk{w}{1};\Uk{w}{2})$ be homogeneous with respect to the first and second derivatives: $\tilde{E}(z;\lambda\Uk{w}{1};\lambda\Uk{w}{2})=\lambda^k\tilde{E}(z;\Uk{w}{1};\Uk{w}{2})$.
Solutions of the system of equations \eqref{e19} are the invariants
\begin{equation}\label{e20}
J_i=w_i,\quad 1\leq i\leq n;\quad J_{ij}=w_i^2 w_{jj}+w_j^2 w_{ii}-2w_iw_jw_{ij},\quad 1\leq i<j\leq n.
\end{equation}
Thus, up to a positive constant factor, the second-order differential equations invariant under the transformations \eqref{e17} can be represent in the form
\begin{equation}\label{e21}
 \tilde{E}(z;\Uk{w}{1};\Uk{w}{2})=F(z,\tilde{J}_i,\tilde{J}_{ij})=0,
 \end{equation} 
where $\tilde{J}_i=w_i/w_n;\ \tilde{J}_{ij}=J_{ij}/w_n^3,\ F(\cdot,\cdot,\cdot)$ is an arbitrary function.

Let $z^n=u$ and $x^a=z^a,\ (a=1,\dots,n-1)$ be the dependent and independent variables, respectively. 
Subsequently,
\[
u_a=-\tilde{J}_a,\quad u_{aa}=-\tilde{J}_{an},\quad u_{ab}=(\tilde{J}_{ab}-\tilde{J}_{a}^2\tilde{J}_{bn} - \tilde{J}_{b}^2\tilde{J}_{an})/2\tilde{J}_{a}\tilde{J}_{b}.
\]
These expressions for the derivatives $u_a, u_{ab}$ in terms of the invariants $\tilde{J}_{a},\tilde{J}_{ab}$ allow us to rewrite an arbitrary differential equation \eqref{e13} in the form of  \eqref{e21}. $\hfill\triangleright$

To conclude this section, we formulate, without giving any proofs, the following theorem.

\begin{theorem}\label{t2}
A differential equation (a system of differential equations) of any order with dependent variables $u=(u^1,\dots,u^m)$ can be represented in the covariant form,
\begin{eqnarray}\label{e22}
&&E_\mu(z;w,\Uk{w}{1},\dots,\Uk{w}{r_{\mu}})=0,\quad \mu=1,\dots s;\\ \label{e23}
&&w^{\alpha}=0,\quad \alpha=1,\dots,m.
\end{eqnarray}
The system of equations \eqref{e22}, \eqref{e23} is invariant under the transformations
\[
w^\alpha\to s^\alpha_\beta(z)w^\beta,\quad \det \| s^\alpha_\beta(z) \|\neq 0.
\]
\end{theorem}

\section{Differential equations of simply transitive type}
\label{sec:3}

Let us consider a second-order partial differential equation of type $\{n,{2},1,n,0,0\}$, i.e., an  equation of the form \eqref{e13} admitting an $n$-dimensional group $G$ of point symmetries with infinitesimal generators~\eqref{e4}.
The group $G$ acts simply transitively on an $n$-dimensional space $Z$ of $n-1$ independent variables $x^a=z^a,\; (a=1,\dots,n-1)$ and one dependent variable $u=z^n$. 
By virtue of the simple transitivity and local character of the problem, we can assume that the infinitesimal generators of the symmetry group form the realization of the corresponding Lie algebra $\g$ by the left-invariant vector fields: $X_i = \xi_i(z) = (L_z)_* e_i$, $i = 1, \dots, n$.
Again, as Section \ref{sec:5}, we denote a basis of the Lie algebra of right-invariant vector fields by $\eta_i = (R_{z^{-1}})_* e_i = \eta_i^j(z) \p_{z^j}$.

For the case being considered here, the dependent variable $\{ u \}$ is not invariant under $G$, and we therefore cannot use Theorem \ref{t3} directly.
Nevertheless, the transition to the covariant form, where the dependent variable $\{ w \}$ is an invariant of the symmetry group $G$, allows us to determine the general form of differential equations with simply transitive symmetry groups.
 
\begin{theorem}\label{t4}
The covariant form of any second-order partial differential equation with a simply transitive group of  point symmetries has the form
\begin{equation}\label{e25}
F(I_{(i)},I_{(ij)})=0,\quad w=0,
\end{equation}
where
\begin{equation}\label{e26}
I_{(i)}=\w{i}/\w{n},\quad I_{(ij)}=(\w{i}^2 \w{jj}-2 \w{i}\w{j}\w{ij}+\w{j}^2 \w{ii})/\w{n}^3,\quad 
1\leq i<j \leq n.
\end{equation}
\end{theorem}

\emph{Proof.} $\triangleleft$
In accordance with Proposition~\ref{u3}, any second-order differential equation~\eqref{e13} can be written in the covariant form~\eqref{e14}, \eqref{e15}, where the dependent variable $w$ is an invariant of the symmetry group $G$. 
Therefore, we can utilize Theorem~\ref{t3} and represent \eqref{e14} in the form \eqref{e24} (for $k=2$). 
By virtue of Theorem \ref{t1}, we also require invariance of \eqref{e24} under the infinite-dimensional group of transformations~\eqref{e17}. 
The second prolongation of the generators of this transformation group can be written as
\begin{multline*}
\Uk{X}{2}=\sigma(z)w\frac{\p}{\p w}+\sum\limits_{i}D_{z^i}(\sigma(z)w)\frac{\p}{\p w_i}+
\sum\limits_{i\leq j}D_{z^i}D_{z^j}(\sigma(z)w)\frac{\p}{\p w_{ij}}=\\ 
=\sigma(z)w\frac{\p}{\p w}+\sum\limits_{i}\hat{\eta}_i(\sigma(z)w)\frac{\p}{\p \w{i}}+
\sum\limits_{i\leq j}(\hat{\eta}_i\circ\hat{\eta}_j)(\sigma(z)w)\frac{\p}{\p \w{ij}}.
\end{multline*}
Then, considering \eqref{e15}, we have
\[
\Uk{X}{2}F(\Uk{w}{(1)},\Uk{w}{(2)})|_{F=0} = \left(\sigma(z) D + \sum\limits_j\sigma_{(j)}(z) R_{(j)}\right)F(\Uk{w}{(1)},\Uk{w}{(2)})|_{F=0}=0.
\]
Here, $\Uk{w}{(1)}=\{\w{i}\},\ \Uk{w}{(2)}=\{\w{ij}\},\ \sigma_{(j)}(z)=\hat{\eta}_j\sigma(z)$, and
\[ 
D=\sum\limits_{i} \w{i}\frac{\p}{\p \w{i}}+\sum\limits_{i\leq j}\w{ij}\frac{\p}{\p \w{ij}},\quad
R_{(j)}=\sum\limits_i(1+\delta_{ij})\w{i}\frac{\p}{\p \w{ij}}
\]
By solving the system of differential equations, similar to \eqref{e18} and \eqref{e19}, we obtain  \eqref{e25} and~\eqref{e26}.~$\hfill\triangleright$

\section{Second-order differential equations admitting simply transitive symmetry groups of low dimensions}
\label{sec:4}

Let $\g^n$ be the Lie algebra of an $n$-dimensional local simply transitive symmetry Lie group $G$.
Let us provide an algorithm for calculating the second-order differential invariants of the symmetry group from the structure constants $C_{ij}^k$ of the Lie algebra $\g^n$.
This algorithm also allows us to output the representatives of the classes of differential equations invariant under the symmetry group and linear in the highest order derivatives.
Any other equation admitting the Lie algebra $\g^n$ as a symmetry algebra can be reduced to one of the given representatives by changing the dependent and independent variables.

\begin{description}
\item[Step 1.] Given the structure constants of the Lie algebra $\g^n$, left- $\{\xi_i=\xi_i^k(z)\p_{z^k}\}$ and right-invariant $\{\eta_i=\eta_i^k(z)\p_{z^k}\}$ vector fields are constructed by applying the method proposed in~\cite{MagMikShi15}. 
\item[Step 2.] Introducing the notation $x^a=z^a \ (a=1,\dots,n-1),\ u=z^n$, the infinitesimal generators of the symmetry group are written as $X_i=\xi_i^a(x,u)\p_{x^a}+\xi_i^n(x,u)\p_{u}$.
\item[Step 3.] In accordance with \eqref{ed1}, $\w{i},\;\w{ij}$ are expressed as functions of variables $(x,u,\Uk{w}{1},\Uk{w}{2})$.
\item[Step 4.] Using \eqref{e26}, the invariants $\I{a},\;\I{ij}$ are calculated as functions of the variables $(x,u,\Uk{w}{1},\Uk{w}{2})$.
\item[Step 5.] From the formulae for the implicit derivatives \eqref{e12} and \eqref{e16}, the inverse equalities are derived: 
\begin{equation}\label{epod}
w_a=-u_a w_n,\quad w_{ab}=-w_n u_{ab}-w_{bn}u_a-w_{an}u_b-u_au_bw_{nn}.
\end{equation}
Substituting these into the invariants $\I{a},\;\I{ij}$, we obtain the first- and second-order differential invariants: 
$$
v_a=v_a(x,u,\Uk{u}{1}), \quad v_{ij}=v_{ij}(x,u,\Uk{u}{1},\Uk{u}{2}).
$$
(The invariants $\I{a},\;\I{ij}$ are arranged so that the ``unnecessary'' variables $\{w_n,w_{an},w_{nn}\}$ cancel and are not presented in the invariants $v_a,\;v_{ij}$).

\item[Step 6.] The invariant equation linear in the second derivatives is constructed: 
\begin{equation}\label{einv_eq}
E(x,u,\Uk{u}{1},\Uk{u}{2})=\sum\limits_{1\leq i< j \leq n}a^{ij}(\Uk{v}{1})v_{ij}+b(\Uk{v}{1})=0.
\end{equation}
\item[Step 7.] The following results are summarized in a table: the non-zero commutation relations of the Lie algebra, $\g^n$, the infinitesimal generators $X_i$ of the symmetry group, a basis of the first- and second-order differential invariants $v_a,\,v_{ij}$, and the invariant differential equation linear in the second derivatives $E(x,u,\Uk{u}{1},\Uk{u}{2})=0$.
\end{description}

As an illustration of the algorithm, we calculate the second order differential invariants and write down the general form of the invariant differential equation linear in the second derivatives for the three-dimensional Lie algebra $\mathfrak{so}(3)$:
\begin{equation*}
[e_1, e_2] = e_3,\quad
[e_2, e_3] = e_1,\quad
[e_1, e_3] = - e_2.
\end{equation*}
Before making this, we calculate the left- and right-invariant vector fields on the corresponding local Lie group. 
In the coordinates of the second kind $g_z = e^{z^3 e_3}e^{z^2 e_2}e^{z^1 e_1}$, these can be constructed by applying the algebraic method described in~\cite{MagMikShi15}:
\begin{equation*}
\xi_1 = \p_{z^1},\
\xi_2 = \sin {z^1} \tan z^2 \p_{z^1} + \cos z^1 \p_{z^2} + \frac{\sin z^1}{\cos z^2}\, \p_{z^3},\
\xi_3 = \cos {z^1} \tan z^2 \p_{z^1} - \sin z^1 \p_{z^2} + \frac{\cos z^1}{\cos z^2}\, \p_{z^3};
\end{equation*}
\begin{equation*}
\eta_1 = \frac{\cos z^3}{\cos z^2}\, \p_{z^1} - \sin z^3 \p_{z^2} + \cos z^3 \tan z^2 \p_{z^3},\ 
\eta_2 = \frac{\sin z^3}{\cos z^2}\, \p_{z^1} + \cos z^3 \p_{z^2} + \sin z^3 \tan z^2 \p_{z^3},\ 
\eta_3 = \p_{z^3}.
\end{equation*}

The infinitesimal generators $X_i$ of the symmetry group are obtained  from the left-invariant vector fields~$\xi_i$ by the formal substitute $z^1 \to x, z^2 \to y, z^3 \to u$:
\begin{equation*}
X_1 = \p_{x},\quad
X_2 = \sin {x} \tan y\, \p_{x} + \cos x\, \p_{y} + \frac{\sin x}{\cos y}\, \p_{u},\quad
X_3 = \cos {x} \tan y\, \p_{x} - \sin x\, \p_{y} + \frac{\cos x}{\cos y}\, \p_{u}.
\end{equation*}
The quantities $w_{(i)}$ and $w_{(ij)}$, which are defined by \eqref{ed1} and considered as functions of the variables $x, y, u, w_1, w_2, w_{11}, w_{12}, w_{13}, w_{22}, w_{23}, w_{33}$, take the form
\begin{equation*}
w_{(1)} = w_1\, \frac{\cos u}{\cos y} - w_2 \sin u + w_3 \cos u  \tan y,\quad
w_{(2)} = w_1\, \frac{\sin u}{\cos y} + w_2 \cos u + w_3 \sin u  \tan y,\quad
w_{(3)} = w_3,
\end{equation*}
\begin{multline*}
w_{(11)} = w_{11}\, \frac{\cos^2 u}{\cos^2 y} - w_{12}\, \frac{\sin 2u}{\cos y} + w_{13}\, \frac{2 \tan y \cos^2 u}{\cos y} + w_{22} \sin^2 u - w_{23} \sin 2 u \tan y +
\\
+ w_{33} \tan^2 y \cos^2 u - w_1\,\frac{\sin 2 u \tan y}{\cos y} - w_2 \cos^2 u \tan y - \frac{1}{2}\, w_3 (1 + 2 \tan^2 y) \sin 2 u,
\end{multline*}
\begin{multline*}
w_{(12)} = w_{11}\,\frac{\sin 2u}{2 \cos^2 y} + w_{12}\,\frac{\cos 2 u}{\cos y} + w_{13}\,\frac{\sin 2 u \tan y}{\cos y} - \frac{1}{2} \, w_{22} \sin 2 u + w_{23} \tan y \cos 2 u + 
\\
+ \frac{1}{2}\, w_{33} \sin 2u \tan^2 y +  w_1\, \frac{\tan y \cos 2 u}{\cos y} - \frac{1}{2}\, w_2 \sin 2u \tan y + \frac{1}{2}\,w_3 \left ( 1 + 2 \tan^2 y \right ) \cos 2 u,
\end{multline*}
\begin{equation*}
w_{(13)} = w_{13}\, \frac{\cos u}{\cos y} - w_{23} \sin u + w_{33} \cos u \tan y - \frac{1}{2}\, w_1\,\frac{\sin u}{\cos y} - \frac{1}{2} \, w_2 \cos u - \frac{1}{2}\, w_3 \sin u \tan y,
\end{equation*}
\begin{multline*}
w_{(22)} = w_{11}\,\frac{\sin^2 u}{\cos^2 y} + w_{12}\,\frac{\sin 2 u}{\cos y} + 2 w_{13}\,\frac{\tan y \sin^2 u}{\cos y} + w_{22} \cos^2 u + w_{23} \sin 2 u \tan y + 
\\
+ \frac{1}{2}\, w_{33} \left ( 1 - \cos 2 u \right ) \tan^2 y + w_1\,\frac{\sin 2 u \tan y}{\cos y} - w_2 \sin^2 u \tan y + \frac{1}{2}\, w_3 \sin 2 u \left ( 1 + 2 \tan^2 y \right ),
\end{multline*}
\begin{equation*}
w_{(23)} = w_{13}\,\frac{\sin u}{\cos y} + w_{23} \cos u + w_{33} \sin u \tan y + \frac12\, w_1\,\frac{\cos u}{\cos y} - \frac{1}{2}\, w_2 \sin u + \frac{1}{2}\, w_3 \cos u \tan y,
\end{equation*}
\begin{equation*}
w_{(33)} = w_{33}.
\end{equation*}

Using the expressions for $w_{(a)}$ and $w_{(ij)}$, we obtain the invariants $I_{(a)}$ and $I_{(ij)}$ in accordance with \eqref{e26}. 
Substituting \eqref{epod} into the invariants, we find
\begin{equation}
\label{1inv}
v_{1} = u_x \, \frac{\cos u}{\cos y} - u_y \sin u - \cos u \tan y,\quad
v_{2} = u_x\, \frac{\sin u}{\cos y} + u_y \cos u - \sin u \tan y,
\end{equation}
\begin{multline*}
v_{12} = \frac{1}{\cos^2 y} \left [ 
u_{xx} u_y^2 + 2 u_{xy} u_y (\sin y - u_x) + u_{yy} (u_x - \sin y)^2 + \frac{1}{2} \, ( 1 - u_y^2  ) u_y \sin 2 y - \right .
\\
\left . - 3 u_x u_y \cos y - 2 (1+u_x^2) u_y \tan y + \frac{4 u_x u_y}{\cos y} \right ],
\end{multline*}
\begin{multline*}
v_{13} = u_{xx}\,\frac{\cos^2 u}{\cos^2 y} - u_{xy}\,\frac{\sin 2u}{\cos y} + u_{yy} \sin^2 u + \frac{1}{2} \left ( u_y^2 - \frac{u_x^2}{\cos^2 y} \right ) \sin 2 u - u_x u_y \, \frac{\cos 2 u}{\cos y}  - 
\\
- u_y \tan y \sin^2 u + \frac{\sin 2 u}{1 + \cos 2 y},
\end{multline*}
\begin{multline*}
v_{23} = u_{xx} \, \frac{\sin^2 u}{\cos^2 y} + u_{xy} \, \frac{\sin 2 u}{\cos y} + u_{yy} \cos^2 u - \frac{1}{2} \left ( \frac{1-u_x^2}{\cos^2 y} + u_y^2 \right ) \sin 2 u +
\\
+ u_x u_y\,\frac{\cos 2 u}{\cos y} - u_y \cos^2 u \tan y.
\end{multline*}
Instead $v_{12}$, $v_{13}$, and $v_{23}$, it is convenient to choose the new set of invariants $\tilde{v}_{12}$, $\tilde{v}_{13}$, $\tilde{v}_{23}$, which is related to the old one by the relations
\begin{equation*}
\tilde{v}_{12} = \frac{v_{12}-v^2_{1}v_{23}-v^2_{2}v_{13}}{v_{1}v_{2}},\quad
\tilde{v}_{13} = v_{13} - v_{23},\quad
\tilde{v}_{23} = v_{13} + v_{23}.
\end{equation*}
As a result, we have
\begin{equation*}
\tilde{v}_{12} = \frac{u_{xx}}{\cos^2 y} + u_{yy} - u_y \tan y,
\end{equation*}
\begin{equation*}
\tilde{v}_{13} = \cos 2 u \left ( \frac{u_{xx}}{\cos^2 y} - u_{yy} - \frac{2 u_x u_y}{\cos y} + u_y \tan y \right ) + \sin 2 u \left ( - \frac{2 u_{xy}}{\cos y} + u_y^2 + \frac{1-u_x^2}{\cos^2 y} \right ),
\end{equation*}
\begin{equation*}
\tilde{v}_{23} = - \sin 2 u \left ( \frac{u_{xx}}{\cos^2 y} - u_{yy} - \frac{2 u_x u_y}{\cos y} + u_y \tan y \right ) + \cos 2 u \left ( - \frac{2 u_{xy}}{\cos y} + u_y^2 + \frac{1-u_x^2}{\cos^2 y} \right ),
\end{equation*}

In accordance with \eqref{einv_eq}, the general form of the $\mathfrak{so}(3)$-invariant differential equation linear in the second derivatives can be written as
\begin{multline*}
\left [ a^{13}(v_{1},v_{2}) \cos 2 u - a^{23}(v_{1},v_{2}) \sin 2 u \right ] \left ( \frac{u_{xx}}{\cos^2 y} - u_{yy} - \frac{2 u_x u_y}{\cos y} + u_y \tan y \right ) +
\\
+ \left [ a^{13}(v_{1},v_{2}) \sin 2 u + a^{23}(v_{1},v_{2}) \cos 2 u \right ] \left ( - \frac{2 u_{xy}}{\cos y} + u_y^2 + \frac{1-u_x^2}{\cos^2 y} \right ) +
\\
+ a^{12}(v_{1},v_{2}) \left ( \frac{u_{xx}}{\cos^2 y} + u_{yy} - u_y \tan y \right ) + b(v_{1},v_{2}) = 0,
\end{multline*}
where $a^{12}, a^{13}, a^{23}$, $b$ are arbitrary smooth functions of their arguments, and $v_{1}$ and $v_{2}$ are defined by~\eqref{1inv}.
We can yet further simplify this equation by dividing by $a^{12}$ and re-designating the rest arbitrary functions:
\begin{multline}
\label{so3eq}
\Delta u + \left [ \tilde{a}^{13}(v_{1},v_{2}) \cos 2 u - \tilde{a}^{23}(v_{1},v_{2}) \sin 2 u \right ] \left ( \frac{u_{xx}}{\cos^2 y} - u_{yy} - \frac{2 u_x u_y}{\cos y} + u_y \tan y \right ) +
\\
+ \left [ \tilde{a}^{13}(v_{1},v_{2}) \sin 2 u + \tilde{a}^{23}(v_{1},v_{2}) \cos 2 u \right ] \left ( - \frac{2 u_{xy}}{\cos y} + u_y^2 + \frac{1-u_x^2}{\cos^2 y} \right ) + \tilde{b}(v_{1},v_{2}) = 0.
\end{multline}
Here, $\tilde{a}^{13} = a^{13}/a^{12}$, $\tilde{a}^{23} = a^{23}/a^{12}$, $\tilde{b} = b/a^{12}$, $\Delta$ is the Laplace--Beltrami operator on the 2-sphere:
\begin{equation*}
\Delta u \equiv \frac{u_{xx}}{\cos^2 y} + u_{yy} - u_y \tan y.
\end{equation*}
It is interesting to note that if the functions $\tilde{a}^{13}$ and $\tilde{a}^{23}$ are constants, then the coefficients of $u_{xx}$, $u_{xy}$, and $u_{yy}$ in \eqref{so3eq} do not depend on $u_x$ and $u_y$. 
In particular, if $\tilde{a}^{13} = \tilde{a}^{23} = 0$, then the coefficients do not also depend on  $u$ and can be regarded as the components of the standard metric on the 2-sphere $S^2$.
In this case, the differential equation \eqref{so3eq} has the form: 
\begin{equation*}
\Delta u + \tilde{b} \left ( v_{1}, v_{2} \right ) = 0.
\end{equation*}

Below, we list the second-order differential invariants and the invariant quasi-linear second-order differential equations for all two- and three-dimensional local Lie groups.
To denote the Lie algebras, we use the notation $\g^n_k$, where $n$ is the dimension of a Lie algebra and $k$ is its number in the list.
In addition to the invariants and differential equations, the tables also contain simply transitive realizations of the Lie algebras in the ``splitting'' coordinates $z = (x,u)$.

\subsection{Two-dimensional symmetry groups (type $\{2,\{2\},1,2,0,0\}$)}

As we have noted, in the two-dimensional case, there exist only two non-isomorphic Lie algebras: the abelian Lie algebra $2 \g^1 \simeq \mathbb{R}^2$ and the non-abelian Lie algebra $\g^2$ isomorphic to the Lie algebra of the affine group $\mathrm{Aff}(\mathbb{R})$.

\begin{table}[!h]
\centering
\small
\caption{\small Second-order differential invariants of two-dimensional symmtry groups and quasi-linear differential equations of type $\{2,\{2\},1,2,0,0\}$}\label{tab:2}
\vspace{4mm}
\begin{tabular}{|l|l|l|l|}
\hline 
Lie algebra & Generators $X_i$ & Invariants  & Quasi-linear second-order 	\\
	&	&	$v_{a}$, $v_{ij}$	&		 differential equation	 \\ 
\hline 
$2\g^1$ & $X_1=\p_x$, $X_2=\p_u$	&	$u_x$, $u_{xx}$ & $u_{xx}+b(u_x)=0$ \\ 
\hline
$\g^2$ & $X_1=\p_x$, $X_2=x\p_x+\p_u$ &	$e^u u_x$, $e^{2u}u_{xx}$& $u_{xx}+e^{-2u}b(e^u u_x)=0$ \\ 
$[e_1,e_2]=e_1$	&	&	&	\\
\hline 
\end{tabular} 
\end{table}

In Table \ref{tab:2}, the symbol $b$ denotes an arbitrary smooth function of its argument.

\subsection{Three-dimensional symmetry groups (type $\{3,\{2\},1,3,0,0\}$)}

We use the classification of non-isomorphic three-dimensional Lie algebras suggested by Mu\-ba\-rak\-zya\-nov~\cite{Mub63}.

\begin{center}
\small
\LTcapwidth=\textwidth
\begin{longtable}{|l|l|l|l|}
\caption{\small Second-order differential invariants of three-dimensional symmetry groups and quasi-linear differential equations of type $\{3,\{2\},1,3,0,0\}$.}\label{tab:3}\\
\hline
Lie algebra	&	Generators $X_i$	& 	Invariants 	&	Quasi-linear second-order differential	 \\
	&	&	$v_a,\ v_{ij}$	&	 equation	\\
\hline
$3\g^1$	&	$X_1=\p_x,\ X_2=\p_y$, &	$u_x, u_y, u_{xx}$, 	&	
$u_{xx} + a_1(u_x,u_y) u_{xy} + a_{2}(u_x, u_y) u_{yy} + $	
\\
	&	$X_3 = \p_u$	&	$u_{xy}, u_{yy}$	&	$\hfill + b(u_x, u_y) = 0$	\\
\hline
$\g^1+\g^2$ 	&	$X_1=\p_x$, &	$e^y u_x, u_y$, &	$u_{xx} + e^{-y}a_1(e^y u_x,u_y) u_{xy} + \hfill$	\\
$[e_1,e_2]=e_1$	&	$X_2=x\p_x+\p_y$,	&	$e^{2y} u_{xx}, e^y u_{xy}$,	&	$\hfill+e^{-2y} a_{2}(e^yu_x, u_y) u_{yy} + \hfill$	\\
	&	$X_3 = \p_u$	&	 $u_{yy}$	&	$\hfill+e^{-2y} b(e^y u_x, u_y) = 0$	\\
\hline
$\g^3_1$ &	$X_1 = \p_u$, $X_2 = \p_x$, &	$u_x - y$, $u_y$, 	& $u_{xx} + a_{1}(u_x - y,u_y) u_{xy} + \hfill$	\\
$[e_2,e_3]=e_1$	&	$X_3 = \p_y + x \p_u$	& $u_{xx}$, $u_{xy}$, $u_{yy}$	& $\hfill+a_{2}(u_x-y, u_y) u_{yy} +b(u_x-y,u_y) = 0$	\\
\hline
$\g^3_2$ 	&	$X_1 = \p_u$, $X_2 = \p_y$, &	$e^{-x} u_x$, $u_y - x$, 	&	
$e^{-x}u_{xx} + a_{1}(e^{-x} u_x, u_y - x) u_{xy} + \hfill$	\\
$[e_1,e_3]=e_1$ &	$X_3 = \p_x + y \p_y + \hfill$	&	$e^{-x} u_{xx}$, $u_{xy}$, 	&	$\hfill + a_{2}(e^{-x}u_x,u_y-x) u_{yy} + \hfill$	\\
$[e_2,e_3]=e_1+e_2$	&	$\hfill+(y+u) \p_u$	&	$e^x u_{yy}$	&	$\hfill + b(e^{-x}u_x,u_y - x) = 0$	\\
\hline	
$\g^3_3$ &	$X_1=\p_x$, $X_2=\p_y$, &	$e^u u_x$, $e^u u_y$, &	
$u_{xx} + a_1(e^u u_x, e^u u_y) + \hfill$	\\
$[e_1,e_3]=e_1$ &	$X_3 = x\p_x +  \hfill$ &	$ e^{2u} u_{xx}$, $e^{2u}u_{xy}$, &	$\hfill+a_2(e^u u_x, e^u u_y) u_{yy} + \hfill $	\\
$[e_2,e_3]=e_2$	&	$\hfill + y\p_y + \p_u$	&	$e^{2u}u_{yy}$	&	$\hfill+e^{-2u} b(e^u u_x, e^u u_y) = 0$	\\
\hline
$\g^3_4$ &	$X_1=\p_x$, $X_2=\p_y$, &	$e^u u_x$, $e^{hu} u_y$, &	$u_{xx} + e^{(h-1)u}a_1(e^u u_x, e^{hu} u_y) + \hfill$	\\
$[e_1,e_3]=e_1$ &	$X_3 = x\p_x+\hfill$	&	$\ e^{2u}u_{xx}$, &	$\hfill+e^{2(h-1)u}a_2(e^u u_x, e^{hu} u_y) u_{yy} + \hfill$	\\
$[e_2,e_3]=he_2$&	$\hfill+hy\p_y+\p_u$	&	$e^{(1+h) u}u_{xy}$, &	$\hfill+e^{-2u} b(e^u u_x, e^{hu} u_y) = 0$	\\
$|h|\leq 1,$ $h \neq 0,1$	&	&	$e^{2hu}u_{yy}$	&	\\
	&	&	&	\\
\hline
$\g^3_5$ &	$X_1 = \p_x$, $X_2 = \p_y$, &	$v^5_1$, $v^5_2$, &	$u_{xx} + u_{yy} + e^{-2pu} b(v^5_1, v^5_2) + \hfill$		\\
$[e_1,e_3]=pe_1-e_2$ &	$X_3 = (px+y)\p_x+\hfill$	&	$v^5_{12}$, $v^5_{13}$, $v^5_{23}$	&	$\hfill+a_1(v^5_1, v^5_2) (u_{xx} - u_{yy}) \cos 2 u - \hfill$	\\
$[e_2,e_3]=pe_2+e_1$ &	$\hfill+(py-x)\p_y+\p_u$	&	&	$\hfill-2 a_1(v^5_1, v^5_2) u_{xy} \sin 2 u + \hfill$	\\
$p\geq 0$	&	&	&	$\hfill+a_2(v^5_1, v^5_2) (u_{xx} - u_{yy}) \sin 2 u + \hfill$	\\
	&	&	&	$\hfill+2 a_2(v^5_1, v^5_2) u_{xy} \cos 2 u = 0$	\\
\hline
$\g_{3}^6$ 	&	$X_1 = \p_x$, &	$v^6_1$, $v^6_2$, &	
$e^{2y} u_{xx} + 4 u e^{y} u_{xy} + 4 u^2 u_{yy} + \hfill$ \\
$[e_1,e_2]=e_1$ 	&	$X_2 = x \p_x + \p_y$, &	$v^6_{12}$, $v^6_{13}$, $v^6_{23}$	&	$\hfill+2 e^y u_x u_y + 4 u u_y^2 - 4 u^2 u_y+ \hfill$	\\
$[e_1,e_3]=2e_2$ 	&	$X_3 = x^2 \p_x + \hfill$	&	&	$ + a_1\left ( v^6_1, v^6_2 \right ) \left ( e^y u_{xy} + 2 u u_{yy} - u^2 \right ) + \hfill$	\\
$[e_2,e_3]=e_3$		&	$\hfill+2 x \p_y +e^y \p_u$	&	&	$\hfill+a_2 \left ( v^6_1, v^6_2 \right ) \left ( u_{yy} - u \right ) + b\left ( v^6_1, v^6_2 \right ) = 0$	\\
\hline
$\g^3_7$ &	$X_1 = \p_x$, &	&	$\frac{u_{xx}}{\cos^2 y} + u_{yy} - u_y \tan y + b\left (v^7_1,v^7_2\right ) + \hfill$	\\	
$[e_1,e_2] = e_3$	&	$X_2 = \sin x \tan y \p_x + \hfill$	&	$v^7_1$, $v^7_2$,	&	$+\left [ a_1\left (v^7_1,v^7_2\right ) \cos 2 u - a_2\left (v^7_1,v^7_2\right ) \sin 2 u \right ] \times$	\\
$[e_2,e_3]=e_1$	&	$\hfill+\cos x \p_y + \frac{\sin x}{\cos y} \p_u,$	&	$v^7_{12}$, $v^7_{13}$, $v^7_{23}$	&	$\times \left ( \frac{u_{xx}}{\cos^2 y} - u_{yy} - \frac{2 u_x u_y}{\cos y} + u_y \tan y \right ) +$	\\
$[e_3,e_1]=e_2$	&	$X_3 = \cos x \tan y \p_x  - \hfill$	&	&	$ + \left [ a_1 \left ( v^7_1,v^7_2 \right ) \sin 2 u + a_2\left (v^7_1, v^7_2\right ) \cos 2 u \right ]\times$	\\
	&	$\hfill-\sin x \p_y + \frac{\cos x}{\cos y} \p_u$	&	&	$\times\left ( - \frac{2 u_{xy}}{\cos y} + u_y^2 + \frac{1-u_x^2}{\cos^2 y} \right ) = 0$	\\
\hline
\end{longtable}
\end{center}

In Table \ref{tab:3}, the symbols $a_1$, $a_2$, and $b$ denote arbitrary functions of their arguments. 
Furthermore, we use the following notation:
\begin{equation*}
v^5_1 = e^{pu} \left ( u_x \cos u - u_y \sin u \right ),\quad
v^5_2 = e^{pu} \left ( u_x \sin u + u_y \cos u \right ),\quad
v^5_{12} = e^{2pu}(u_{xx}+u_{yy}),
\end{equation*}
\begin{equation*}
v^5_{13} = e^{2pu} \left (  (u_{xx}-u_{yy}) \cos 2 u  - 2 u_{xy} \sin 2 u \right ),\quad
v^5_{23} = e^{2pu} \left (  (u_{xx}-u_{yy}) \sin 2 u  + 2 u_{xy} \sin 2 u \right );
\end{equation*}
\begin{equation*}
v^6_1 = e^y u_x + u_y^2,\quad
v^6_2 = u_y - u,\quad
v^6_{12} = u_{yy} - u,
\end{equation*}
\begin{equation*}
v^6_{13} = e^y u_{xy} + 2 u u_{yy} - u^2,\quad
v^6_{23} = e^{2y} u_{xx} + 2 u \left [ u( 2 u_{yy} + u_y - u ) + e^y ( u_x + 2 u_{xy}) \right ];
\end{equation*}
\begin{equation*}
v^7_1 = u_x\,\frac{\cos u}{\cos y} -  u_y \sin u - \cos u \tan y,\quad
v^7_2 = u_x\,\frac{\sin u}{\cos y} u_x + u_y \cos u - \sin u \tan y,
\end{equation*}
\begin{equation*}
v^7_{12} = \frac{u_{xx}}{\cos^2 y} + u_{yy} -  u_y \tan y,
\end{equation*}
\begin{equation*}
v^7_{13} = \cos 2 u \left ( \frac{u_{xx}}{\cos^2 y} - u_{yy} - \frac{2 u_x u_y}{\cos y} + u_y \tan y \right ) + \sin 2 u \left ( - \frac{2 u_{xy}}{\cos y} + u_y^2 + \frac{1-u_x^2}{\cos^2 y} \right ),
\end{equation*}
\begin{equation*}
v^7_{23} = - \sin 2 u \left ( \frac{u_{xx}}{\cos^2 y} - u_{yy} - \frac{2 u_x u_y}{\cos y} + u_y \tan y \right ) + \cos 2 u \left ( - \frac{2 u_{xy}}{\cos y} + u_y^2 + \frac{1-u_x^2}{\cos^2 y} \right ).
\end{equation*}

\section*{Conclusion}

In this paper, we have offered a systematic procedure for the group classification of differential equations based on the regular actions of their symmetry groups.
Unlike the traditional approaches based on the preliminary group classification, we do not restrict the form of an equation by some ansatz and proceed rather similarly to Lie's original two-step procedure. This procedure involves finding all the inequivalent realizations for a given abstract Lie algebra $\g$  and the subsequent construction of the differential equations that are invariant under each of these realizations.

One of the main advantages of our method is that there is no need to explicitly integrate differential equations. 
This is possible if one considers the subtle features of the actions of the prescribed symmetry groups; in this relation, we have classified all the differential equations into five types, which are listed in Table~\ref{tab:1}.
Another distinguishing feature of our method is its covariance, which is achieved by considering the dependent and independent variables equally.
It allows us to describe differential equations invariant under different transformation groups in a unified manner.
For these purposes, we have introduced the so-called \textit{covariant form} of a differential equation, i.e., equations~\eqref{e22} and \eqref{e23}.
The transition to the typical form of the differential equation suggests the procedure of ``splitting'' of the variables into dependent and independent, and all the equations obtained by different approaches of such ``splitting'' are related by hodograph transformations.

In this study, we have restricted ourselves to scalar second-order differential equations of types $\{n,{2},1,n,0,0\}$ and $\{n+m,\{2\},1,n,0,0\}$ in accordance with Table~\ref{tab:1}.
The first type presupposes a symmetry group acting freely and transitively on the space of dependent and independent variables, whereas the symmetry groups for the equations of the second type are assumed to be free but intransitive, in general.
We have given the most general form of the equations for these types (Theorems \ref{t5} and \ref{t4}) and also formulated constructive algorithms of their group classification.
Using these algorithms, we have obtained  exhaustive classifications of the equations of types $\{n,{2},1,n,0,0\}$ and $\{n+m,\{2\},1,n,0,0\}$ for all the local regular transformation groups of dimensions up to three (see Tables \ref{tab:4}--\ref{tab:3}).

In future investigations, we plan to further develop our classification program; particularly, we will consider in detail the cases of differential equations of types III, IV, and V.    
There are also certain reasons to assume that the algebraic technique we have designed will constructively allow the calculation of operators of the invariant differentiation, which suggests the possibility of effectively constructing differential invariants of any order~\cite{Ovs82, Olv93}.

\bibliographystyle{unsrt}
\bibliography{biblio}

\end{document}